# A Computational Study of Vertical Partial Gate Carbon Nanotube FETs

Youngki Yoon, James Fodor, and Jing Guo*
Department of Electrical and Computer Engineering
University of Florida
Gainesville, FL, 32611

**ABSTRACT**

A vertical partial gate carbon nanotube (CNT) field-effect transistor (FET), which is amenable to the vertical CNT growth process and offers the potential for a parallel CNT array channel, is simulated using a self-consistent atomistic approach. We show that the underlap between the gate and the bottom electrode (required for isolation between electrodes) is advantageous for transistor operation because it suppresses ambipolar conduction. A vertical CNTFET with a gate length that covers only 1/6 of the channel length has a much smaller minimum leakage current than one without underlap, while maintaining comparable on current. Both n-type and p-type transistor operations with balanced performance metrics can be achieved on a single partial gate FET by using proper bias schemes. Even with a gate underlap, it is demonstrated that increasing the CNT diameter still leads to a simultaneous increase of on current and minimum leakage current. Along with a partial gate, the simulated transistor features a significant amount of air between the surface of the channel CNT and the gate insulator, as is caused by the vertical CNT growth process. Filing this pore with a high-κ insulator is shown to have the potential to decrease the on current, due to electrostatic phenomena at the source-channel contact.

*guoj@ufl.edu

**I. Introduction**

Significant progress on potential nanoelectronic applications for carbon nanotube (CNT) has been made in recent years. Most CNT field-effect transistors (FETs) demonstrated to date have a horizontal structure with an individual single wall carbon nanotube (SWNT) channel. Small on current and large parasitic capacitance significantly limit their potential applications. Recent advancements in SWNT fabrication techniques [1-3] have led to the successful implementation of a two-terminal resistive SWNT device [4] within a vertically aligned porous anodic alumina (PAA) template [5]. The SWNTs can be placed within predefined locations in the PAA template using established catalyst methods [6, 7], and good electrical contacts between the SWNTs and metal have also been demonstrated [8]. Coupling this vertical two-terminal PAA-based device with a side gate can potentially lead to a vertical coaxially gated CNTFET with ideal gate electrostatic control [9, 10], or a vertical parallel array CNTFET with large on current and significantly reduced parasitic effects [11].

The effect of non-uniform gate geometry along the channel direction on the horizontal FETs was previously explored experimentally [12] by using a trench, and theoretically [13] in terms of electrostatic engineering via varying gate oxide thickness along the channel. However, a study of the device performance for the vertical partial gated CNTFET has not yet been performed. While some of the physical properties of the vertical CNTFET may be shared by the horizontal CNTFET previously studied, a more careful approach for the vertical CNTFET is necessary to accentuate the key differences between the two devices. A careful study of the vertical CNTFET may help develop central design rules required for the implementation of a successful manufacturing process. In this study, we simulated a vertical partial gate Schottky barrier (SB) CNTFET using self-consistent atomistic simulations. The simulation results indicate that the underlap required for isolation between the gate and the bottom electrodes is advantageous for transistor operation, as it suppresses the ambipolar conduction. Significant increase of the underlap length has only a small effect on the on current and subthreshold swing, but significantly lowers the minimum leakage current. Variations of the partial gate length, therefore, should not lead to a nanotransistor variability problem. Both n-type and p-type transistor operations with balanced performance metrics can be achieved on a single partial gate FET by using proper bias schemes, which is advantageous for CMOS electronics applications. It is demonstrated that increasing the diameter of the channel CNT will increase both the on-current and the minimum leakage current. This is attributed to a smaller band gap, and thus smaller Schottky barrier height. One of the features of the two-terminal PAA SWNT array is a considerable amount of air between the surface of the channel CNT and the PAA template material, which is shown not to have any negative effect on the transistor performance. Within the context of the simulation model, the air pore improves the transistor on current while leaving the off current unaffected.

## II. Approach

*2.1. Device Structure*

We simulated coaxial vertical partial gate SWNT SBFETs at room temperature ($T = 300$ K), with a power supply voltage of $V_{DD} = 0.4$ V. A cross section of the structure and the simulated region is shown in Fig. 1. The nominal device has a 120 nm-long (17, 0) CNT as the channel material, which is overlapped by the gate for 60 nm from the source-channel contact. Two insulators are present between the gate and the CNT—2 nm of PAA and 10 nm of $Al_2O_3$, both of which have an approximate dielectric constant of 10. A 2 nm-radius air pore ($\varepsilon_r = 1$) is treated between the gate oxide and the channel CNT. Both the channel length and air pore/gate oxide radius are scaled down by a factor of 10 with respect to a typical PAA process to facilitate atomistic simulations. Scaling down the device is preferred for efficient gate control and fast switching. The qualitative conclusions on device physics and design rules should apply to larger vertical CNTFETs. The metal source (drain) is directly attached to the CNT channel, and the Schottky barrier height between source (drain) and channel is treated as being equal to half the band gap of the channel material ($\Phi_{Bn} = E_g/2$) [14, 15]. The metal contact Fermi level lies in the middle of the CNT band gap. The CNT is assumed to be at the center of the air pore. In practice, the CNT may be placed near the PAA wall. However, the qualitative conclusions of the work remain the same regardless of the exact position of the CNT in the air pore.

*2.2 Quantum Transport*

The DC characteristics of ballistic CNTFETs are simulated by solving the Schrödinger equation using the non-equilibrium Green's function (NEGF) formalism self-consistently with the Poisson equation. A tight binding (TB) Hamiltonian with a $p_z$ orbital basis set is used to describe an atomistic physical observation of the channel. A $p_z$ orbital coupling parameter of 3 eV is used and only the nearest neighbor coupling is considered. The mode space approach is utilized to achieve significant computational cost savings. Only the first lowest subband is considered for carrier transport, and all the CNTs simulated are intrinsic.

The retarded Green's function of the CNT channel is given by

$$\mathbf{G^r} = \left[(E+i0^+)\mathbf{I} - \mathbf{H} - \mathbf{\Sigma_1} - \mathbf{\Sigma_2}\right]^{-1}, \qquad (1)$$

where $\mathbf{H}$ is TB Hamiltonian matrix of the CNT channel with a $p_z$ orbital basis set, $\mathbf{\Sigma_{1,2}}$ is self-energy of the metal source (drain) contact.

The charge density can be computed as,

$$Q(x) = (-q)\int_{-\infty}^{+\infty} dE \cdot \text{sgn}[E - E_N(x)]\{D_1(E,x)f\left(\text{sgn}[E - E_N(x)](E - E_{F1})\right)$$

$$+ D_2(E,x)f\left(\text{sgn}[E - E_N(x)](E - E_{F2})\right)\}, \quad (2)$$

where $q$ is the electron charge, $\text{sgn}(E)$ is the sign function, $E_{F1,F2}$ is the source (drain) Fermi level, and $D_{1,2}(E,x)$ is the local density of states due to the source (drain) contact, which is computed by the NEGF method. The charge neutrality level, $E_N(x)$, is at the middle of band gap because the conduction band and the valence band of the CNT are symmetric.

The source-drain current is calculated as

$$I_D = \frac{4q}{h}\int_{-\infty}^{\infty} \text{Trace}\left[\mathbf{\Gamma}_1 \mathbf{G}^r \mathbf{\Gamma}_2 \mathbf{G}^{r+}\right](f_1 - f_2)dE, \quad (3)$$

where $\mathbf{\Gamma}_{1,2} = i\left(\mathbf{\Sigma}_{1,2} - \mathbf{\Sigma}_{1,2}^+\right)$ is broadening function of the source (drain) contact, $f_{1,2}$ is equilibrium Fermi function of source (drain) contact, and the factor of 4 comes from spin and valley degeneracies of two [16]. The gate leakage current is neglected for simplicity. The atomistic simulation approach used in this study has been carefully validated by experimental measurements [17], and it shows a very good quantitative agreement.

*2.3 Electrostatics*

The self-consistent potential is computed from the charge density and the electrode potentials using the Poisson equation,

$$\nabla \cdot \left[\varepsilon(\vec{r})\nabla U(\vec{r})\right] = qQ(\vec{r}) \quad (4)$$

where $U(\vec{r})$ is the electron potential energy which determines the diagonal entry of the potential energy matrix in Eq. (1), $\varepsilon(\vec{r})$ is the permittivity, and $Q(\vec{r})$ is the charge density. Because the electric field only varies in 2D dimensions of r and z axes for the simulated coaxial device structure, the simulated region is simply chosen from a cross section of the device structure as shown by the dashed line in Fig.1a. The Poisson equation is numerically solved using the finite element method (FEM) because it naturally lends itself to treating complex device geometries as well as boundaries between different dielectric materials.

**III. RESULTS**

*3.1 $I_D$-$V_G$ characteristics*

Fig. 2a shows $I_D$-$V_G$ plot at $V_D$ = -0.4 V. Compared to the $I_D$-$V_G$ of a full gate CNT SBFET, which has symmetric ambipolar characteristics, vertical partial gate CNT SBFETs show nearly unipolar characteristics. Fig. 2b and 2c show the conduction and valence band profiles along the channel position for $V_G$ = -0.9 V and $V_G$ = 0.6 V, respectively. At $V_G$ = -0.9 V, the electrostatic potential of the channel at the source end is sufficiently controlled by the gate electrode, and the Schottky barrier is thin enough for holes to tunnel though the barrier. However, at $V_G$ = 0.6 V, the channel potential at the drain end is not effectively controlled by the gate electrode, and hence electrons in the drain encounter a very thick Schottky barrier and hardly flow through the channel due to significant gate underlap at the drain end. This partial gate effect is advantageous for transistor operation because it suppresses the ambipolar conduction of typical full gate CNT SBFETs, thus resulting in smaller minimum current and a larger maximum on-off current ratio.

*3.2 Device operation*

Partial gate CNT SBFETs allow either p-type or an n-type transistor operation, provided that the terminals are correctly biased. Note that the bottom electrode, which is not overlapped by the gate, should be used as the drain contact. The top electrode, which is overlapped by the gate, is appropriately used as the source for both n-type and p-type operation. The source is grounded ($V_S$ = 0 V) for both types of transistor operation. For p-type operation, the drain must have a negative voltage, with a negative gate voltage used to turn on the transistor. This biasing causes a small Schottky barrier between the source and channel, which allows holes to flow into the channel. Fig. 3a shows the on state of p-type operation at $V_G = V_D = -V_{DD}$, where $V_{DD}$ is the power supply voltage. Similarly, for n-type operation, the drain voltage must be positive, with a positive gate voltage required for operation in the on state. This biasing also causes a reduction in the Schottky barrier between source and channel, in this case enabling electron conduction through the channel. Fig. 3b shows the on state of n-type operation at $V_G = V_D = V_{DD}$. The insets shown in Fig. 3a and 3b are schematic drawings of the bias schemes for p-type and n-type operation, respectively. In both cases the key parameter that determines the current flow through the channel is the source-channel Schottky barrier, as the partial gate can only modulate the electrostatic potential at this end. Restated, the gate control over the channel electrostatic potential on the drain side is so poor that the Schottky barrier between the drain and channel cannot be effectively reduced with the partial gate. The Schottky barrier between the drain and channel always remains thick, regardless of gate voltage. This is demonstrated in Fig. 2.

*3.3 $I_D$-$V_D$ characteristics*

Fig. 4a shows the $I_D$ versus $V_D$ characteristics of a partial gate CNT SBFET with gate voltages of $V_G$ = -0.5, -0.6, and -0.7 V. For a full gate CNT SBFET it is known that source-drain current is linearly increased as drain voltage is increased, provided that the gate voltage is sufficiently high. This is not the case for a partial gate CNT SBFET. When the drain voltage is in the range of $-0.1 < V_D < 0$ V, the carriers injected from the source cannot reach the drain because of the thick barrier between the drain and the channel, which causes severe reflection of carriers. This is a direct result of the poor control of the electrostatic potential at the drain side [13]. Fig. 4b shows the band profile along the channel position for $V_D = -0.1$ V (the solid line) and $V_D = -0.4$ V (the dashed line). As the applied drain voltage increases, the barrier at the drain end is reduced, which enables carriers to be collected by the drain. The source-drain current exponentially increases with $V_D$ because the reflection probability of the source-injected carriers by the drain barrier exponentially decreases as $V_D$ increases. As the drain voltage further increases, the drain-side Schottky barrier vanishes and saturation current is achieved (the potential profile at the source end remains unchanged because the electrostatic potential near the source is totally controlled by the gate voltage).

*3.4 The effect of gate length*

As mentioned previously, using a partial gate suppresses the ambipolar characteristic present in a full gate CNT SBFET. It is useful to explore how the gate length (or equivalently the length of the underlap between the gate and the drain electrodes) affects the device performance of the vertical CNTFET. Fig. 5a shows the $I_D$-$V_G$ curve of partial gate FETs for gate lengths $L_g =$ 20, 60, and 120 nm at $V_D = -0.4$ V. All other parameters are equal to those of the nominal device described in Fig. 1b. The results show that the ambipolar characteristic of full gate CNT SBFETs is further suppressed as the gate length is reduced because the underlap length has a direct effect on the Schottky barrier between drain and channel. At $V_G = 0.6$ V, the source-drain current of 60 nm and 20 nm gate lengths is 4 and 5 orders of magnitude smaller than that of a 120 nm gate length, respectively. It can also be shown that the partial gate geometry highly affects the minimum leakage current, such that a much smaller minimum leakage current can be obtained by using a shorter gate length. The minimum leakage current associated with a 20 nm gate length is smaller than that of a 120 nm gate length by an order of 2. Fig. 5b shows the conduction and the valence band profiles along the channel position for different gate lengths at $V_G = 0.6$ V. Even though the partial gate has a severe effect on the minimum leakage current, it has a weak effect on the on-current of the device. This is because the gate voltage has good control over the Schottky barrier between the source and the channel, which completely determines the carrier injection from the source. At $V_G = -0.6$ V, the difference of source-drain currents between $L_g = 60$ nm and 120 nm is less than 5%. The subthreshold swing is also almost invariant to the gate length.

*3.5 The effect of air pore*

In the nominal device, it is assumed that there is an air pore surrounding the CNT. To explore the consequence of this air pore, a simulation was conducted with the pore filled with the same material as the gate insulator. Fig. 6a shows the resulting $I_D$-$V_G$ curve. The results are interesting in that the on current for the device lacking an air pore is smaller than that of the nominal device structure. This result can be understood via careful observation of the Schottky barrier between the source and channel. Fig. 6b compares the electric field pattern for both cases at a bias of $V_D = -0.4$ V and $V_G = -0.6$ V. It demonstrates that the redirection of the electric field caused by the air/insulator interface actually reduces the barrier between the source and channel. Near the source contact, the electric field is more heavily controlled by the source potential.

Since the dielectric of air is smaller than that of the insulator, the air pore essentially makes it more difficult for the source to modulate the electric field near the source contact. The electric field modulation caused by the source increases the source-channel Schottky barrier, and thus suppressing this modulation decreases the barrier. At $V_G$ = -0.6 V, the source-drain current is reduced by 34% when the air pore is filled with $Al_2O_3$. However, the off current is not affected by filling the pore, since the Schottky barrier at drain side is so thick that the decrement is hard to recognize. At $V_G$ = 0.6 V, the current of both cases does not show any appreciable difference.

*3.6 The effect of CNT diameter*

The effect of CNT diameter is explored by using different chiral index to define the channel CNT. $I_D$-$V_G$ characteristics for CNT channels of various diameters are shown in Fig. 7. As the diameter of the CNT increases, so does the on current and minimum leakage current. Usually, a CNT with a larger diameter has a smaller band gap (this is true for the chiral index ranges of CNTs simulated in this study, although it is invalid for very small diameter tubes due to curvature effects). Therefore, a larger diameter CNT channel has smaller Schottky barrier heights at the source and drain contacts, which results in increases in both on current and minimum leakage current. At a large positive gate voltage, the electron leakage current is limited by the thick Schottky barrier at the drain end with a barrier height of $\Phi_{Bn} = E_g/2$, which decreases as the diameter increases. In general, a larger on current and a smaller minimum leakage current are desired for transistors, so the effect of CNT diameter causes a device performance trade-off. The source-drain current of a (26, 0) CNTFET is a factor of 33 larger than that of a (17, 0) CNTFET at $V_G$ = -0.4 V. At the same time, the minimum leakage current of a (26, 0) CNTFET is on the order of 2 times larger than that of a (17, 0) CNTFET.

**IV. Conclusions**

A computational study on the device performance of vertical partial gate CNT SBFETs was performed using the NEGF formalism self-consistently with the Poisson equation. The features of non-uniform gate geometry along the channel of the horizontal CNTFETs can be generally adopted to the vertical CNTFETs, which essentially has a partial gate for the isolation of electrodes. A more careful approach is required for vertical CNTFETs, however, due to unique fabrication issues. The underlap between the gate and the drain electrode is advantageous for transistor operation since it suppresses the ambipolar conduction of typical full gate CNTFETs. Increasing the underlap length has only small effects on the on current and subthreshold swing, while the minimum leakage current is significantly reduced. Both n-type and p-type transistor operations with balanced performance metrics are possible with a single partial gate CNTFET, provided that proper bias schemes are used. Using a larger diameter CNT channel significantly increases both on current and minimum leakage current. The CNT diameter must be carefully considered due to the strong trade-off it causes in device performance. The results also show a counter-intuitive result in that the air pore between the CNT channel and the gate insulator improves the on current while leaving the off current unaffected.

**ACKNOWLEDGEMENTS**

It is our pleasure to thank Dr. Thomas Tombler of Atomate Corp. and Profs. Timothy Fisher and David Janes of Purdue University for discussions.

**FIGURE CAPTIONS**

Fig. 1. A vertical partial gate CNTFET, which can be fabricated by placing a side insulator and gate to the already demonstrated two-terminal CNT devices on porous anodic alumina (PAA). (a) The cross section of the modeled geometry. The coaxial structure offers the best gate electrostatics and has qualitatively similar device physics to planar gate CNTFETs. (b) The device parameters of the nominal device. Both length and radius are scaled down by a factor of 10 from a typical PAA process to facilitate atomistic simulations, which does not result in any change of qualitative conclusions on device physics. The (17, 0) CNT has a diameter of ~1.3 nm and band gap of 0.63 eV. The insulator oxide is $Al_2O_3$, which has the same dielectric constant as PAA ($\varepsilon_r \approx 10$).

Fig. 2. *Suppression of ambipolar conduction* (a) $I_D$ versus $V_G$ characteristic at $V_D$ = -0.4 V for the vertical CNT SBFET as shown in Fig. 1. (b) The band profile along the channel position at $V_G$ = -0.9 V. The gate has good control over the channel electrostatic potential at the source side and greatly decreases the Schottky barrier, thus allowing easy hole conduction. (c) The band profile at $V_G$ = 0.6 V. The gate has very poor control over the Schottky barrier at the drain end due to the partial gating. Therefore, electrons hardly go through the Schottky barrier from the drain.

Fig. 3. *Device operations* (a) The conduction and valence bands profile along the channel position for p-type operation with $V_G = V_D = -V_{DD} = -0.4$ V, where $V_{DD}$ is the power supply voltage. It is observed that the gate controls the Schottky barrier at source end well and allows hole conduction. (b) The band profile for n-type operation with $V_G = V_D = V_{DD} = 0.4$ V. The gate controls the Schottky barrier at source end, and allows electron flowing.

Fig. 4. (a) $I_D$ versus $V_D$ characteristics of the partial gate CNT SBFET with different gate voltages of $V_G$ = -0.5, -0.6, and -0.7 V. (b) The conduction and the valence band profile along the channel position at $V_D$ = -0.1 V (the solid line) and $V_D$ = -0.4 V (the dashed line) with $V_G$ = -0.5 V.

Fig. 5. *The effect of the gate length* (a) $I_D$ versus $V_G$ curves of partial gate FETs for the partial gate lengths of $L_g$ = 20, 60 and 120 nm at $V_D$ = -0.4 V. The underlap between the gate and the drain decreases as the gate length increases (the channel length is invariant with $L_{ch}$ = 120 nm). It is observed that the device becomes more ambipolar and exhibits a much larger minimum leakage current as the underlap decreases. (b) The band profile along the channel position at $V_G$ = 0.6 V. Dotted line is for 20 nm of gate length, dashed line is for 60 nm, and solid line is for 120 nm. A larger underlap between the gate and the drain suppresses ambipolar characteristic of CNTFET and reduce minimum leakage current.

Fig. 6. (a) $I_D$ versus $V_G$ for the cases of a CNT within an air pore and within a fully filled gate oxide. Filling the air pore with $Al_2O_3$ even decreases the on-current. (b) Electric field contour near the source contact for devices with an air pore of 2 nm thick (left) and without air pore (right) at $V_D$ = -0.4 V and $V_G$ = -0.6 V. The Schottky barrier is thicker in the case of the filled pore, which can be shown in the electric field pattern.

Fig. 7. *$I_D$ versus $V_G$ for CNT channels of various diameters*: Solid line is for $n$ = 17, solid line with circles is for $n$ = 20, dashed line is for $n$ = 23, and dashed line with squares is for $n$ = 26. As the diameter of the CNT increases, so does the on current and minimum leakage current. A larger diameter CNT has a smaller band gap and a smaller Schottky barrier height, which results in larger source-drain current.

**FIGURES**

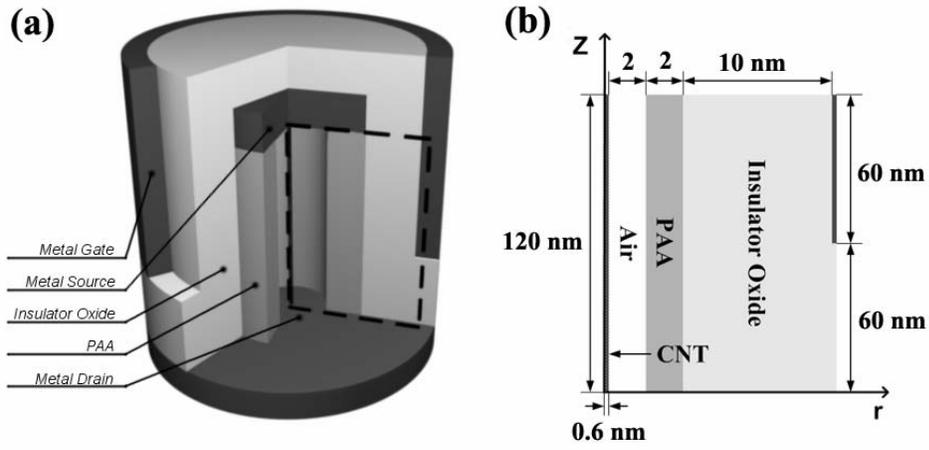

Fig. 1

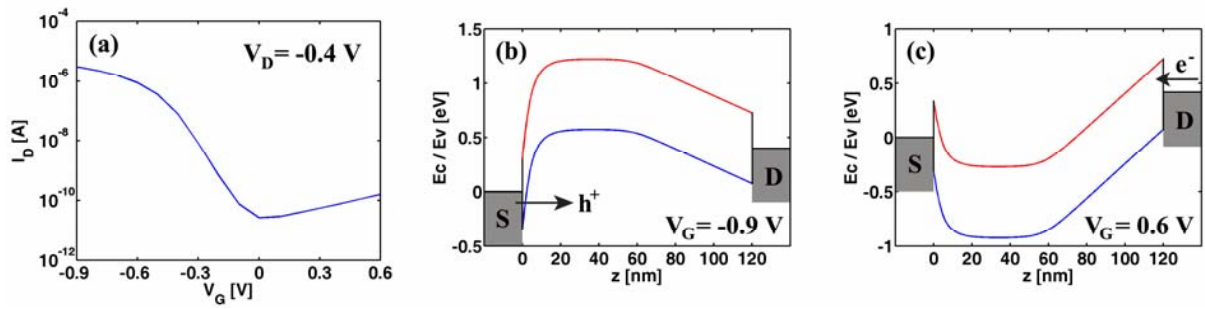

Fig. 2

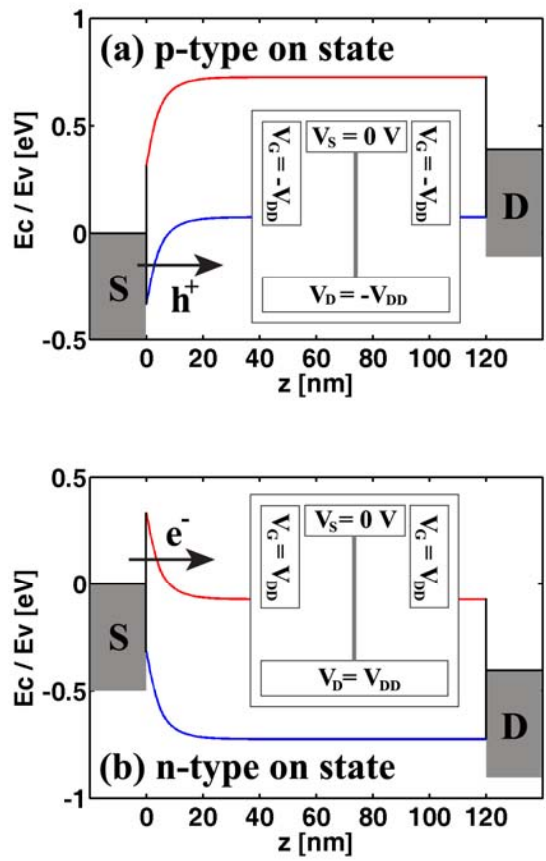

Fig. 3

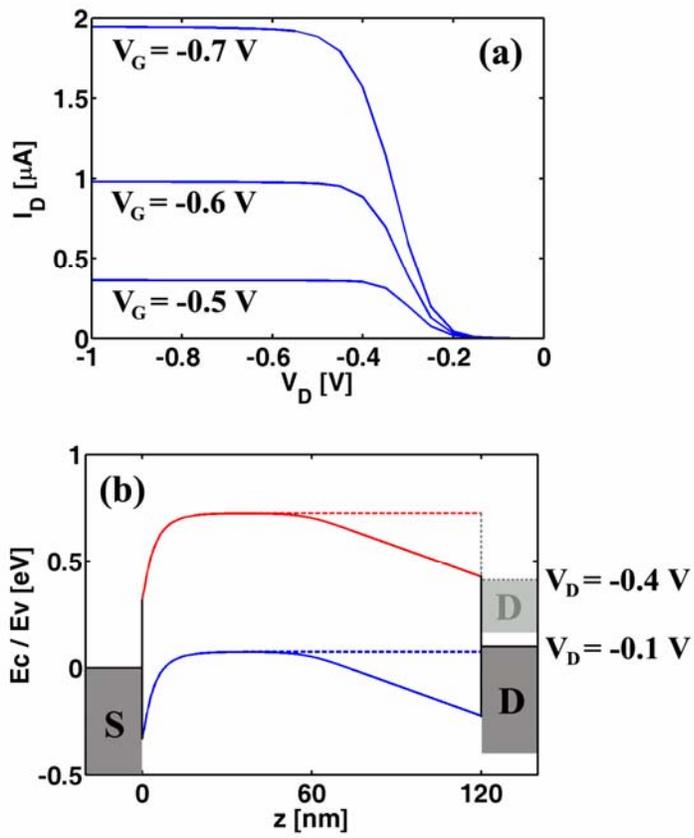

Fig. 4

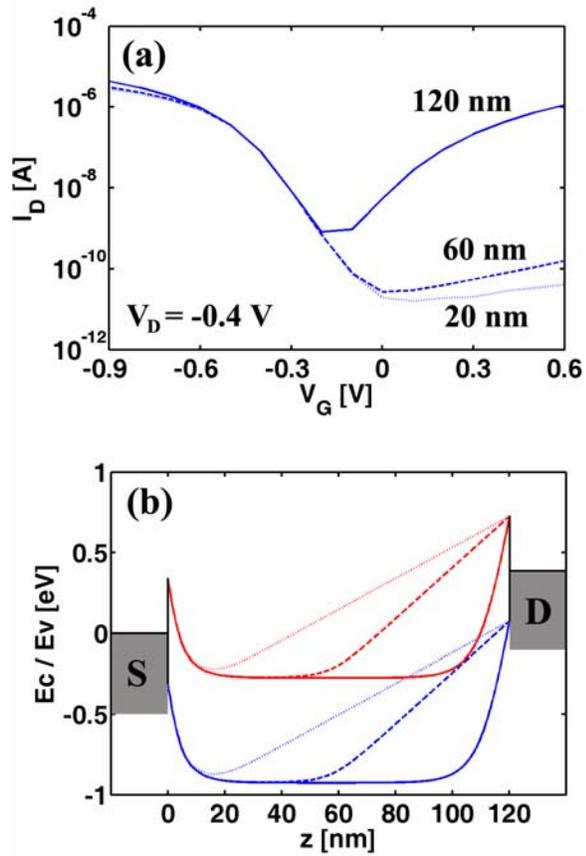

Fig. 5

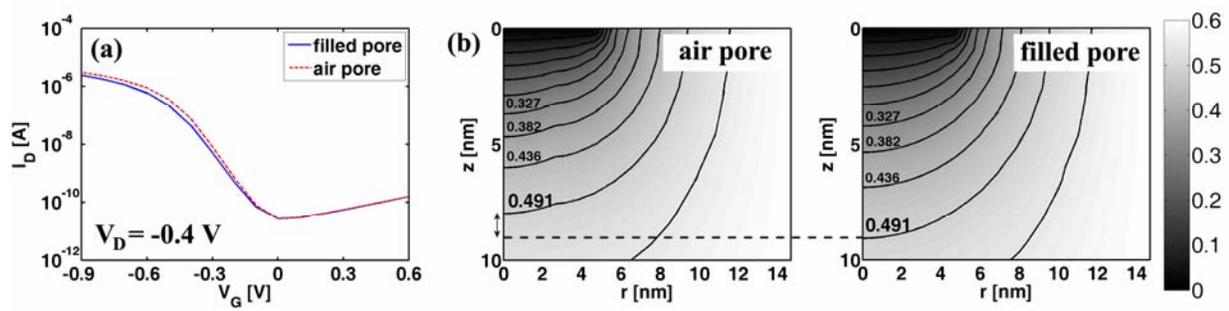

Fig. 6

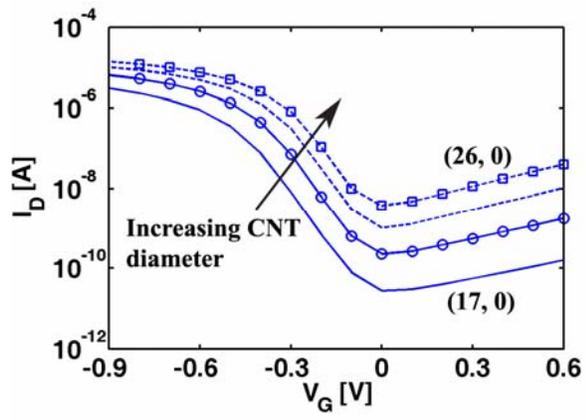

Fig. 7